\documentclass{vldb}
\pdfoutput=1
\usepackage{caption}

\usepackage[ruled,vlined]{algorithm2e}
\usepackage{color}

\newcommand{\rjmnew}[1]{\textcolor{black}{#1}}

\newcommand{\kennew}[1]{\textcolor{black}{#1}}

\newcommand{\eat}[1]{}
\newcommand{\FP}{\mathrm{FP}}
\newcommand{\FN}{\mathrm{FN}}

\newcommand{\NFP}{N^\mathrm{FP}}
\newcommand{\sig}{\mathrm{MinHash}}
\newcommand{\dom}{\mathrm{dom}}
\newcommand{\cost}{\mathrm{cost}}
\newcommand{\Hs}[1]{\Hat{s}_{#1}}
\newcommand{\Ht}[1]{\Hat{t}_{#1}}
\usepackage{tabularx}
\usepackage{ragged2e}  \newcolumntype{Y}{>{\RaggedRight\arraybackslash}X}
\usepackage{booktabs}  \usepackage{textcomp}
\DeclareMathOperator*{\argmin}{arg\,min}
\newtheorem{prop}{Proposition}
\newtheorem{definition}{Definition}
\newtheorem{theorem}{Theorem}

\newcommand{\squishlist}{
 \begin{list}{$\bullet$}
  { \setlength{\itemsep}{0pt}
     \setlength{\parsep}{1pt}
     \setlength{\topsep}{1pt}
     \setlength{\partopsep}{0pt}
     \setlength{\leftmargin}{1em}
     \setlength{\labelwidth}{1em}
     \setlength{\labelsep}{0.5em} } }

\newcommand{\squishend}{
  \end{list}
}

\definecolor{americanrose}{rgb}{1.0, 0.01, 0.24}
\definecolor{airforceblue}{rgb}{0.36, 0.54, 0.66}
\definecolor{ao(english)}{rgb}{0.0, 0.5, 0.0}
\definecolor{ao}{rgb}{0.0, 0.0, 1.0}

\newcommand{\kenpu}[1]{\textcolor{black}{#1}}
\newcommand{\eric}[1]{\textcolor{black}{#1}}
\newcommand{\rmj}[1]{\textcolor{black}{#1}}
\newcommand{\fatemeh}[1]{\textcolor{black}{#1}}

\newcommand{\ericcr}[1]{\textcolor{black}{#1}}
 
\begin{document}

\title{LSH Ensemble:  Internet-Scale Domain Search}

\eat{
\numberofauthors{4} \author{
\alignauthor
       Erkang Zhu\\
       \affaddr{University of Toronto}\\
       \email{\fontsize{10}{12}\selectfont ekzhu@cs.toronto.edu}
\alignauthor
       Fatemeh Nargesian\\
       \affaddr{University of Toronto}\\
       \email{\fontsize{10}{12}\selectfont fnargesian@cs.toronto.edu}
\alignauthor
       \\
       \affaddr{UOIT}\\
       \email{\fontsize{10}{12}\selectfont ken.pu@uoit.ca}
       \alignauthor Ren{\'e}e J. Miller\\
       \affaddr{University of Toronto}\\
       \email{\fontsize{10}{12}\selectfont miller@cs.toronto.edu}
}
}
\author{
  \begin{minipage}{1.0\linewidth}
    \begin{minipage}[b]{0.23\linewidth}
      \centering
Erkang Zhu\\
\affaddr{University of Toronto}\\
\email{\fontsize{10}{12}\selectfont ekzhu@cs.toronto.edu}
    \end{minipage}
    \begin{minipage}[b]{0.26\linewidth}
            \centering
Fatemeh Nargesian\\
\affaddr{University of Toronto} \\
\email{\fontsize{10}{12}\selectfont fnargesian@cs.toronto.edu}
    \end{minipage}\hspace*{-0.4cm}
    \begin{minipage}[b]{0.3\linewidth}
      \centering
Ken Q. Pu\\
\affaddr{UOIT}\\
\email{\fontsize{10}{12}\selectfont ken.pu@uoit.ca}
    \end{minipage}
    \begin{minipage}[b]{0.2\linewidth}
      \centering
Ren\'ee  J. Miller\\
\affaddr{University of Toronto}\\
\email{\fontsize{10}{12}\selectfont miller@cs.toronto.edu}
    \end{minipage}
  \end{minipage}
}

\clearpage

\pagenumbering{arabic}

\maketitle

\begin{abstract}
We study {\it the problem of domain search} where a domain is a set of distinct values from an unspecified universe.  We use 
Jaccard set containment score, defined as $|Q \cap X|/|Q|$, as the measure of relevance of a domain $X$ to a query domain $Q$.  
Our choice of Jaccard set containment over Jaccard similarity as a measure of relevance makes our work particularly suitable for 
searching Open Data and data on the web, as Jaccard similarity is known to have poor performance over sets with large 
differences in their domain sizes.
We demonstrate that the domains found in several real-life Open Data and web data repositories show a power-law distribution over their domain sizes.

We present a new index structure, Locality Sensitive Hashing (LSH) Ensemble, 
that solves the domain search problem using set containment at Internet scale. 
Our index structure and search algorithm cope with the data volume and skew 
by means of data sketches using Minwise Hashing and domain partitioning. 
Our index structure does not assume a prescribed set of data values.  
We construct a cost model that describes the accuracy of 
LSH Ensemble with any given partitioning.  This allows us to formulate the data partitioning for LSH Ensemble as an optimization problem.  
We prove that there exists an {\it optimal} partitioning for any data distribution.  Furthermore, for datasets 
following a power-law distribution, as observed in Open Data and Web data 
corpora, we show that the optimal partitioning can be approximated using equi-depth, making it particularly efficient to use in practice.

We evaluate our algorithm using real data (Canadian Open Data and WDC Web
Tables) containing up over 262 million domains.  The experiments demonstrate
that our index consistently outperforms other leading alternatives in accuracy
and performance.  The improvements are most dramatic for data with large skew in
the domain sizes.  Even at 262 million domains, our index sustains query
performance with under 3 seconds response time.
\end{abstract}
 
\eat{


}
\section{Introduction}
\label{sec:intro}

In the Open Data movement, large volumes of valuable databases are being published on the Web. 
Governments around the world are launching Open Data portals 
(some of which are shown in Table~\ref{tab:open-data}).
The data format is
highly heterogeneous, comprised of a mixture of relational (CSV and
spreadsheet), semi-structured (JSON and XML), graph based  (RDF), and
geo-spatial formats.  
There 
is an increasing number of datasets in which well-structured attributes
(with or without a name) can be identified,
each containing a set of values that we
will call a {\em domain}.

\begin{table}[]
    \centering
    \begin{tabular}{c|c} 
            {\bf Country} & {\bf Number of Datasets} \\ 
                & {\bf (Structured and Semi-Structured)} \\ \hline
        US & 191,695  \\
        UK & 26,153 \\
        Canada & 244,885 \\
        Singapore & 11,992 \\ 
            \end{tabular}
    \caption{Examples of Governmental Open Data as of First Quarter 2016.}
    \label{tab:open-data}
\end{table}

It is not just federal governments that are releasing massive
numbers of datasets. Several projects have extracted tables from HTML
pages~\cite{Cafarella:VLDB:2008, Lehmberg:WWW:2016}.
Cafarella et al.~\cite{Cafarella:VLDB:2008}
extracted 154 million relational tables from Google's general-purpose
web crawl. Lehmberg et al.~\cite{Lehmberg:WWW:2016} have compiled and
published 51 million relational tables extracted from Common Crawl. 
Tables extracted by Lehmberg et al., called Web Data Commons (WDC) Web Tables, are open and accessible to the public 
outside of search engines. These tables provide a common ground for research 
on Data on the Web.

Despite the openness of these datasets, effectively gaining access \rjmnew{to them}
is 
still
challenging for many reasons.

\squishlist
\item
There is lack of schema description \eat{to}in most of the data\-sets. Identifying relevant datasets is exceedingly difficult.
\item
In most cases, the only form of data access is bulk download.  
This means traditional search at the  data level is impossible at the moment.
\item
 Data is hosted on the Web.  While the Web is a robust and open platform, it  provides very low bandwidth for data access.
  Database operations such as sorting, scan and join are nearly impossible
  over the Web.  This makes {\em ad hoc} access of Open Data \eat{impossible}\rjmnew{impractical}.
\squishend

\subsection{Web Joins Using Domain Search}
\label{sec:web-joins}

\vspace{3mm}
\begin{figure}[t]
	\includegraphics[width=\columnwidth,natwidth=1, natheight=1]{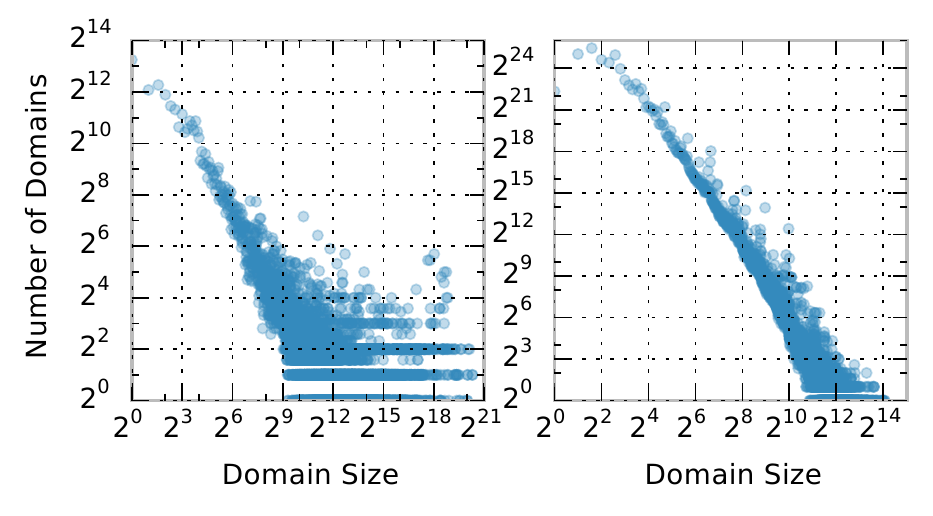}
	\vspace{-3mm}
	\caption{Domain Size Distribution of the Canadian Open Data (Left) and the
	    English Relational Subset of 
	    	    	    WDC Web Table Corpus 2015 (Right).}
	\label{fig:card_dist}
\end{figure}
\vspace{-3mm}

In this paper, we focus on a specific problem of Open Data management: the domain search problem.  
A {\em domain} is simply a set of values.
A dataset can be characterized by one or more domains.  For example, the Canadian federal 
research 
agency 
publishes \rjmnew{information about} the industry partners of successful research grants as relational
tables. One such table {\tt NSERC\_GRANT\_PARTNER\_2011} has the following attributes:

\vspace{1mm}
\noindent
{\small
\texttt{Identifier, Partner, Province, Country, Fiscal Year, ...}
}
\vspace{1mm}

\noindent    
Each attribute contains a domain.   
\rmj{Given such a table, an interesting question is to find other
  tables that join with this table on a specified attribute.  For
  example, to find additional information about industry partners a
  data scientist may wish to find tables that join on
  \texttt{Partner}.  To do this, we must find tables
that include
a domain that {\bf contains} as much of the \texttt{Partner}
  domain as possible.}

We  define the domain search problem as an R-near neighbor problem~\cite{Andoni:CACM:2008}. 
Given a query
domain $Q$ and a relevance threshold $t^*$, find all domains $X$ whose
relevance to $Q$ is within $t^*$.  
The unique characteristics of web data impose 
constraints on a solution to the domain search problem.

\squishlist
\item {\em Scalability}.
    For Internet-scale domain search, a solution must be able to handle hundreds
    of millions of domains.

   \item {\em Open world domains}. 
    A solution cannot assume a fixed vocabulary or prescribed set of values 
    covering all domains.
    As new domains are introduced (potentially with new, unseen values), 
    the search should be able to handle them.
    
    \item {\em Skewed distribution}.
    Some domains are small (a domain of 
    provinces has only a few
        values), 
    while others can be quite large
    (the Canadian government's contract-disclosure data\-set 
            has domains with tens of thousands 
    of distinct values).
            We have collected the Canadian Open Data repository and the English 
    relational subset of 
            WDC Web Tables Corpus 2015~\cite{Lehmberg:WWW:2016}.
    Figure~\ref{fig:card_dist} shows
    the respective domain size distributions.  
    It is clear that in each case, the domain sizes follow a power-law distribution.

    \item {\em Relevance}.
\rmj{For the domain search problem, a domain is relevant if it
  contains as much of the search domain as possible.}
        \rjmnew{Jaccard similarity \rmj{(size of the intersection over the
        size of the union)} is a commonly used measure of relevance
      between sets when the sets have comparable cardinalities.  Open
      data and Web data domains have potentially large differences in
      their domain sizes.  It is known \cite{Agrawal:SIGMOD:2010} that
      Jaccard similarity is not an accurate or intuitive measure of
      relevance due to its bias to smaller domains.   \rmj{Related
        measures, including the cosine similarity or dice coefficient,
        in fact any measure that is a ratio of two functions of both
        domains ($\frac{f(x,y)}{g(x,y)}$) will have the same bias when
        applied over sets with a large skew in their cardinality}.
      For the purpose of domain search, \rmj{where our goal is to find
        domains that contain as much of the query domain as possible,}
      it is more intuitive to define the relevant domains as the ones
      with {\em large} overlap with the query domain.  Thus we choose
      the (Jaccard) {\em set containment} score as the relevance
      measure \rmj{(the size of the intersection divided by the size of the
        query domain)}.}

    \eat{Because of the large skew in domain sizes, some commonly used relevance scores, like the
    {\bf Jaccard similarity} (the size of the intersection of two
    sets over their union) cannot be used~\cite{Agrawal:SIGMOD:2010}.
    For data with skewed domain sizes, {\bf set containment} 
    (the size of the intersection 
            over the size of the query) can be used
    instead~\cite{Agrawal:SIGMOD:2010}.}
    
\item {\em Compact Index Size and Small Query Memory Footprint.}
\kenpu{
    In order to scale to Internet scale of hundreds of millions of domains, 
    we need the index to be highly compact in size.  We also want the
    representation of the search query domain to have small
    memory footprint as it needs to be exchanged over the Web.
    If we use raw values of the domains, the data exchange would degrade
    the query performance significantly due to the relatively low bandwidth of
    the Internet.
}
\squishend

The scalability and compact index size limitations suggest 
using small, fixed size data sketches 
to effectively approximate a
relevance score.
The requirement for supporting
open-world domains makes a hashing approach like the {\em Minwise Hashing}~\cite{Broder:1997} data sketch, or simply MinHash, a natural 
choice to represent domains.
It is known that the Jaccard similarity 
can be accurately estimated using MinHash.
Furthermore, {\em Locality Sensitive Hashing}
(LSH)
is an efficient index 
for approximate R-near neighbor queries~\cite{Andoni:CACM:2008},
and it can be used for the Jaccard similarity search~\cite{Indyk:STOC:1998}.

However, Jaccard similarity does not agree with set containment, and
the disagreement is exasperated when the sizes of the sets being compared
differ significantly.  The most recent approach to MinHash-based indexing that
supports set containment is {\em Asymmetric Minwise
Hashing}~\cite{Shrivastava:WWW:2015}.  This approach ``pads'' the domains with
fresh values so they end up with equal sizes.  \rjmnew{Shrivastava and Li show}
that near neighbors based on Jaccard similarity and padded MinHash signatures
converge
to near neighbors with set containtment~\cite{Shrivastava:WWW:2015}.
We experimentally show here that padding \rjmnew{used on Open Data may} 
decrease recall.  As we will show in Section~\ref{sec:experiment}, with a finite
number of hash functions, increased skewness in the domain sizes may lead to a
significant decrease in recall.
\rjmnew{LSH, Asymmetric Minwise Hashing, and other related work is presented in greater detail in Section~\ref{sec:relatedwork}.}

\subsection{Contributions}

\vspace{1mm}
The main contributions of this paper are as follows.

\squishlist
\item \rmj{We define the domain search problem using set containment as the
  relevance measure so that we can find domains that maximally contain
  a query domain.   We use domain search to find joinable 
tables.  }

\item We present a new indexing structure, the {\em LSH Ensemble}, to
  address the domain search problem for datasets found in Open Data
  and Web data repositories at an Internet scale.  LSH Ensemble is an
  efficient and scalable solution to the domain search problem. 

\item  We propose a data partitioning scheme 
\rmj{that makes LSH Ensemble accurate over domains whose 
sizes are
skewed (following a power-law distribution).}
    
\item 
We present a cost model 
for precision and recall of any partitioning,
enabling us to formulate data
partitioning for LSH Ensemble as an optimization  problem. 
    
\item 
We prove the existence of an optimal partitioning for any data distribution.  
In particular, for power-law distributions (which we have empirically observed in Canadian Open Data and the WDC Web Tables Corpus 2015), 
we show that the optimal partitioning can be approximated using equi-depth, allowing for efficient implementation.

\item
Experimentally, we evaluate our implementation of LSH Ensemble using both the Canadian Open Data repository and WDC Web Tables Corpus 2015.  We show that 
LSH Ensemble scales to \ericcr{hundreds of millions of} domains with each domain having up to millions of distinct values, while consistently sustaining query response time of \ericcr{a few seconds.}

\item
\kennew{We 
compare our approach
against the state-of-the-art 
MinHash LSH~\cite{Indyk:STOC:1998} and Asymmetric 
Minwise Hashing~\cite{Shrivastava:WWW:2015}.  Compared to \eat{other}\rjmnew{these} alternatives, our approach significantly improves 
precision while maintaining high 
recall.  
}
\squishend

\vspace{1mm}
We believe that LSH Ensemble is an essential tool that will enable
data scientists to find new, relevant Open Data and Web Data for their
data analysis projects.   \rmj{By finding domains that maximally
  contain a query dataset, our approach can help scientists to find
  datasets that best augment their own data (by maximally joining with
  their data).}

 \section{Problem Definition}
\label{sec:probdef}

\kenpu{
By {\em domain}, we mean a set of data values.
A data set $R$  can be characterized by a collection of {\em domains}.
When $R$ is a relational table, the domains are given by
the projections $\pi_i(R)$ on each of the attributes of $R$.
We write $\dom(R)$ to denote the set of all domains of $R$.
}

\kenpu{
    In Section~\ref{sec:web-joins}, our motivation is, given a source data set
    $R$, we want to discover data sets which are joinable with $R$.
    We assume that, given a data set $W$, $R$ and $W$ are highly joinable
    if there is a pair of domains $X\in\dom(R)$ and $Y\in\dom(W)$ such
    that $X$ is mostly contained by $Y$.  Namely $|X\cap Y|/|X|$ is large.
}
\begin{definition} [Set containment]
The set containment \newline of 
$X$ in $Y$ is defined as:
\begin{equation}
\label{eq:containment}
    t(X, Y) = \frac{|X \cap Y|}{|X|}
\end{equation}
\end{definition}

\rmj{We formalize the {\em domain search problem} as a $R$-nearest neighbor problem.}
\begin{definition} [Domain Search]
Given a collection of domains $\mathcal{D}$, a query domain $Q$, and a threshold
$t^* \in [0,1]$ on the set containment score, find a set of relevant domains
from $\mathcal{D}$ defined as
\begin{equation}
    \{X : t(Q, X) \ge t^*, X \in \mathcal{D}\}
\end{equation}
\end{definition}

In addition to set containment, one can also quantify the relevance of two domains $X$ and $Y$ using
{\em Jaccard similarity} of the respective domains~\cite{Broder:1997}.
The Jaccard 
similarity between domains 
$X$ and $Y$ is defined as:
\begin{equation}
\label{eq:resemblance}
    s(X, Y) = \frac{|X \cap Y|}{|X \cup Y|}
\end{equation}

While Jaccard similarity is the right measure for many applications, 
we argue that set containment, which is asymmetric, is better suited for domain search. 
Consider for example a query domain \texttt{Q}, 
and two domains, \texttt{Provinces} and \texttt{Locations}, as follows: 
\begin{verbatim}
            Q = {Ontario, Toronto}
    Provinces = {Alberta, Ontario, Manitoba}
    Locations = {Illinois, Chicago, New York City
                 New York, Nova Scotia, Halifax, 
                 California, San Francisco, Seattle,
                 Washington, Ontario, Toronto}
\end{verbatim}
Is \texttt{Q} closer to \texttt{Provinces} or to \texttt{Locations}?
The Jaccard similarity of \texttt{Q} and \texttt{Provinces} is $0.25$, 
while that of \texttt{Q} and \texttt{Locations} is $0.083$.
According to these scores, domain \texttt{Q} is more relevant to domain 
\texttt{Provinces} than \texttt{Locations}, which is a bit counter 
intuitive if we look at the data values.
On the other hand, the set containment of \texttt{Q} and \texttt{Provinces} 
is $0.5$ while that of \texttt{Q} and \texttt{Locations} is $1.0$. 
According to these set containment scores, domain \texttt{Q} is 
judged to be more
relevant to \texttt{Locations} than \texttt{Provinces}.
This example shows that Jaccard similarity favors domains with small size
(\texttt{Provinces}).  Such bias is undesirable for the purpose of discovery of
joinable Web data.  Set containment, on the contrary, is agnostic to the
difference in the sizes of the domains.

\kenpu{
    The domain search problem distinguishes itself from the traditional Web
    keyword search in that the query itself is an entire
    domain with arbitrarily large cardinality.  It would be impractical to treat
    each value in the query domain as a keyword.  Thus, we are interested in a
    search algorithm with approximately constant time query complexity.
    In addition, as discussed in 
the introduction, the domain
    containment search problem has the following constraints.
}

\vspace{1mm}
\textbullet \hspace{0.25mm} $|\mathcal{D}|$ is \eat{in}\rjmnew{on} the order of
hundreds millions.

\vspace{1mm}
\textbullet \hspace{0.25mm} $\{|X|: X\in\mathcal{D}\}$ is very skewed.

\vspace{1mm}
\textbullet \hspace{0.25mm} $|Q|$ can be arbitrarily large.

\vspace{1mm}
Given such constraints, we look for an \textbf{approximate} solution 
to the domain containment search problem, with the following properties.

\vspace{1mm}
\textbullet \hspace{0.25mm} {\em Efficient indexing}: 
        The space complexity of the index structure should be {\em
          \eric{almost} constant} 
\rmj{(growing very sublinearly)}
        with respect to the domain sizes, and linear with respect to the
        number of domains. An index structure can be built in a single-pass
        over the data values of the domains.

\vspace{1mm}
\textbullet \hspace{0.25mm} {\em Efficient search}: The search time complexity should be constant
        with respect to the query domain size and sub-linear with respect
        to the number of domains in $\mathcal{D}$.
\vspace{1mm}

An alternative way of formulating the problem is by defining it as searching for
the top-{\it k} relevant domains.
Since we are interested in discovering domains that can be used in value-based joins, 
search by containment threshold is more natural.
The top-$k$ relevant domains are not guaranteed to have a sufficient level of overlap 
with the given query domain.  
We remark that the two formulations (top-$k$ versus threshold) are closely related and
complementary~\cite{Ilyas:2008:STK}.
 \section{preliminaries}
\label{sec:preliminaries}
Our solution to the domain containment search problem is built on the foundation of 
Minwise Hashing~\cite{Broder:1997} and 
LSH~\cite{Indyk:STOC:1998}.
We present a brief overview of these two techniques.

\subsection{Minwise Hashing}
\label{sec:MinHash}
Broder~\cite{Broder:1997} proposed a technique for estimating the Jaccard similarity 
between domains of any sizes.
In this technique, a domain is converted into a {\it MinHash signature} using 
a set of {\it minwise hash functions}.
For each minwise hash function, its hash value is obtained by
using an independently generated hash function that maps all domain values to 
integer hash values and returns the minimum hash value it observed.

Let $h_\mathrm{min}$ be one such hash function and the minimum hash value of a domain
$X$ be $h_\mathrm{min}(X)$ and $Y$ be $h_\mathrm{min}(Y)$.
Broder showed that the probability of the two minimum hash values being equal is the Jaccard similarity of $X$ and $Y$:
\begin{equation}
	P(h_\mathrm{min}(X) = h_\mathrm{min}(Y)) = s(X, Y) 
\end{equation}
Thus, given the signatures of $X$ and $Y$, we can obtain an unbiased 
estimate of the Jaccard similarity by counting the number of {\it collisions} 
in the corresponding minimum hash values and divide that by the total number 
of hash values in a single signature.

\subsection{Locality Sensitive Hashing}
\label{sec:lsh}
\rjmnew{The} Locality Sensitive Hashing (LSH) index was developed for general approximate 
nearest neighbor search problem in high-dimensional spaces~\cite{Indyk:STOC:1998}.
\rjmnew{LSH} can be applied to the approximate R-near neighbor search problem~\cite{Andoni:CACM:2008}.
An LSH index requires a family of LSH functions.
An LSH function is a hash function whose collision probability is high for inputs that
are close, and low for inputs that are far-apart. The distance measure must be
symmetric. 
A formal definition of LSH functions can be found in the work of Indyk
and Motwani~\cite{Indyk:STOC:1998}.

The minwise hash function $h_\mathrm{min}$ belongs to the family of 
LSH functions for Jaccard distance, as the probability of collision is equal to the 
Jaccard similarity.
For simplicity, we call the LSH index for Jaccard similarity {\it MinHash LSH}.

Next we explain how 
to build a MinHash LSH index.
Assume we have a collection of MinHash signatures of domains generated 
using the same set of minwise hash functions.
The LSH divides each signature into $b$ ``bands'' of size $r$. 
For the $i$-th band, 
a function $H_i = (h_\mathrm{min,1},h_\mathrm{min,2},\dots,h_\mathrm{min,r})$ 
is defined, which outputs the concatenation of the minimum hash values in that band, 
namely, the values $h_\mathrm{min,1}$ to $h_\mathrm{min,r}$.
The function $H_i$ maps a band in a signature to a bucket, so that signatures
that agree on band $i$ are mapped to the same bucket.

Given the signature of a query domain,  LSH maps the signature to 
buckets using the same functions $H_1, \dots, H_b$. 
The domains whose signatures map to at least one bucket where 
the query signature is mapped are the {\it candidates}.
These candidates are returned as the query result.
Hence, the search time complexity of the LSH only depends on the number of minwise
hash functions (or the signature size) and is sub-linear with respect to
the number of domains indexed.

The probability of a domain being a candidate is a function of 
the Jaccard similarity $s$ between the query and the domain.
Given the parameters $b$ and $r$, the function is:
\begin{equation}
	\label{eq:lshprob}
	P(s|b,r) = 1 - (1 - s^{r})^{b}
\end{equation}
Given a threshold on Jaccard similarity, we want the domains that
meet the threshold to have high probability of becoming candidates,
while those do not meet the threshold to have low probability.
This can be achieved by adjusting the parameters $b$ and $r$.

 \section{Related Work}
\label{sec:relatedwork}
In this section, we provide a brief survey of previous work 
related to domain search from different perspectives.

\noindent
\textbf{IR Indexes for Keyword Search.}
Indexes for searching structured datasets given a user-specified keyword 
query have been extensively 
studied~\cite{coffman2014empirical, hristidis2002discover,
  kargar2014meanks, Pimplikar:2012}. 
This work is similar to domain search as search results should 
contain the keywords in the query.
However, in domain search, 
the query itself 
is a domain containing (possibly) large number of data values.
\kennew{
To use keyword search to solve domain search, 
we could
generate a keyword
for each element in the domain. Such a solution can only 
be used for
small \ericcr{domains}.
We
need to handle very large domains containing millions of distinct values. In our
solution, the index search time complexity is constant time with respect to the
query domain size.}
\eric{In addition, the relevance measures in keyword search engines,
  such as the tf-idf score, 
	consider the frequency of each keyword in all 
	documents. This is different from the relevance measure in domain search, which
	considers the degree of overlap between the query and the indexed domain.
	}
\eat{
\fatemeh{Web search join 
engines~\cite{Lehmberg:2015,Yakout:2012:infogather}. 
apply a combination of attribute label and instance-based techniques 
to find attributes with overlapping information. 
LSH Ensemble can be used as a component in web search join engines as 
it searches for highly overlapping attributes for a query domain in
Web scale. 
Note that other web join approaches do not compute containment over
the hundreds of millions of domains that we consider.
}
}

\eat{The processing cost for a query domain can be orders of magnitude larger than a typical
keyword query, thus, requiring different approaches to indexing and query processing.}
\noindent
\textbf{Other LSH Indexes.}
Various LSH indexes have been developed for distance metrics other than Jaccard, 
including 
Euclidean~\cite{Datar:SGG:2004}, Hamming~\cite{Indyk:STOC:1998}, and Cosine distance~\cite{Charikar:STOC:2002}, among others.
These approaches are widely used in 
areas such as image search and document deduplication.
Many such LSH variations require the data to be vectors with fixed dimensions.
While it is possible to convert domains into binary vectors, where each unique data value
is a dimension, binary vector representation is not practical for domain search at Internet-scale.
These indices require prior knowledge of the values in the query domain, \rjmnew{hence,}
they \rjmnew{would} need to be rebuilt whenever a domain with unseen values is inserted.
Thus, LSH indexes, such as  MinHash 
LSH,
that 
do not require a fixed set of domain values are 
better suited for Internet-scale domain search.

SimHash is an LSH index for Cosine distance~\cite{Charikar:STOC:2002},
and it is applicable to domain search, since it does not prescribe a fixed set of
values for all domains.
However, it \eat{is}\rjmnew{has been} shown that MinHash LSH is more computationally efficient than 
SimHash~\cite{Shrivastava:AISTATS:2014}.
Hence, we use MinHash LSH in our experimental evaluation (Section~\ref{sec:experiment}).

\eat{
\paragraph*{Domain search using MinHash LSH}
Duan et al.~\cite{Duan:ISWC:2012} uses MinHash LSH to tackle the problem of instance-based ontology 
matching.
The problem is very close to domain search as each type or class in ontology 
is represented by a set of values.
We compare LSH Ensemble with MinHash LSH experimentally in Section~\ref{sec:experiment},
and show that LSH Ensemble provides better precision and overall accuracy.
}

\noindent
\textbf{Asymmetric Minwise Hashing.}
Asymmetric Minwise Hashing by Shrivastava and Li is the state-of-the-art technique for containment search over
documents~\cite{Shrivastava:WWW:2015}. 
Asymmetric Minwise Hashing applies a pre-processing step called the asymmetric transformation,
that injects fresh padding values into the domains to be indexed.\footnote{For efficiency, the padding is done to the MinHash signatures of the domains rather than to the domains themselves.}
This makes all domains in the index have the same size as the largest 
domain.
They show that Jaccard similarity between an untransformed \rjmnew{signature} (\rjmnew{for} the query) and a transformed \rjmnew{signature} 
(\rjmnew{for} index domain) is monotonic with respect to 
set containment. 
Thus, MinHash LSH can be used to index the transformed domains, such that the domains
with higher set containment scores will have higher probabilities of becoming candidates.
While this technique is very useful for containment search over 
documents such as emails and news articles, 
we observed that the asymmetric transformation reduces recall 
when the domain size distribution is very skewed.
Since Asymmetric Minwise Hashing is the only technique that directly addresses containment search,
we 
include it in our
experimental comparison in Section~\ref{sec:experiment}.

\noindent
\textbf{Schema and Ontology Matching Techniques.}
Examples of unsupervised instance-based schema matching techniques 
are COMA++~\cite{Aumueller:2005} and DUMAS~\cite{BilkeN:DUMAS:2005}. 
These approaches rely on overlap or similarity of instance values.  
In order to make instance-based matchers scalable, 
Duan et al.~\cite{Duan:ISWC:2012} use MinHash LSH and random projection (simhash) 
to do type matching 
based on Jaccard and Cosine similarity. 
The goal of LSH Ensemble is to match a query domain against 
extremely large
number of domains with domain sizes varying from a 
few to millions. LSH Ensemble can be applied in large-scale 
schema and ontology matching, \rjmnew{but to the best of our knowledge, no schema matchers scale to millions of attributes}.

\noindent
\textbf{Methods for Table Search.}
Table search is the problem of finding join candidates in a repository of
relational tables. 
Semantic-based table search approaches enrich tables 
with \eat{ontological}\rjmnew{ontology-based} semantic annotations~\cite{Venetis:VLDB:2011,Limaye:2010:ASW}.
Semantic-based table search is complementary to our approach which uses set containment.
Some table search approaches rely on table context and metadata 
in addition to table content to do 
search~\cite{Lehmberg:2015,Yakout:2012:infogather}. 
\ericcr{
For instance, InfoGather uses 
a matching measure consisting of features such as the similarity of 
text around tables in Web pages, the similarity of each table to 
its own context, the similarity of the URL of the web page containing 
each table, and the number of overlapping tuples in 
tables~\cite{Yakout:2012:infogather}.
Open datasets often lack text context, description and any schema related 
information. Our approach relies solely on the values of domains in tackling
the domain search problem.}
\rmj{LSH Ensemble can be used as a component in 
		table search 
	engines as 
	it searches for highly overlapping attributes to a query domain in
	Web scale. 
		Note that other 
		table search
	approaches are not able to do containment
	search over hundreds of millions of domains as in LSH Ensemble.}
 \begin{table}[]
        \centering
        \caption{Symbols Used in Analysis}
        \label{tbl:symbols}
        \begin{tabularx}{\columnwidth}{@{} c Y c @{}}
        \toprule
        Symbol & Description \\
        \midrule
        $X$, $Q$ & Indexed domain, Query domain \\ \addlinespace
        $x$, $q$ & Domain size of $X$, Domain size of $Q$ \\ \addlinespace
        $b$, $r$ & Number of bands in LSH, Number of hash functions in each band \\ \addlinespace
                $s(Q,X)$, $t(Q,X)$ & Jaccard similarity and containment score between sets $Q$
        and $X$\\ \addlinespace
        $s^*$, $t^*$ & Jaccard similarity and containment thresholds \\ \addlinespace
        $\Hs{x,q}(t)$, $\Ht{x,q}(s)$ & functions that convert $t(Q,X)$ to $s(Q,X)$ and vice
        versa, given the domain sizes $x=|X|$ and $q=|Q|$ \\ \addlinespace
        $\FP$, $\FN$ & False positive \& negative probabilities \\ \addlinespace
                $\NFP$ & Count of false positive domains \\ \addlinespace
        $l$, $u$ & Lower, Upper bound of a partition \\ \addlinespace
        $m$ & Num.~\eat{of minwise}hash functions in MinHash \\ \addlinespace
        $n$ & Num.~of partitions in LSH Ensemble \\ \addlinespace
        \bottomrule
        \end{tabularx}
\end{table}

\section{LSH Ensemble}
\label{sec:contribution}

\eat{In this section, we describe our contribution, {\em LSH Ensemble}.
By partitioning the domains by their sizes, we build an ensemble of MinHash
LSH indexes. For each MinHash LSH index, we compute a conservative Jaccard
similarity threshold based on the containment threshold, so that the resulting
false negatives are minimized.
We further construct a cost model describing the query performance.
Using the cost model, we provide an approximate solution to the optimal partitioning, 
and a query-time tuning of each MinHash LSH index.
The LSH ensemble index provides accurate and performant domain search using containment, even
when the distribution of domain sizes is skewed.
}

\kenpu{
    In this section, we describe our contribution, {\em LSH Ensemble}.
    In our approach, the domains are indexed in {\em two stages}.  In the first
    stage, the domains are partitioned into disjoint partitions based on the
    domain cardinality.
    In the second stage, we construct a MinHash LSH index for {\em each}
    partition.  Each LSH index is dynamically tuned with its specific Jaccard
    similarity threshold.
    In Section~\ref{sec:conservative} to Section~\ref{sec:estimatefp},
    we relate 
set
    containment to Jaccard similarity, and 
present
a cost model that describes the resulting false positive
rate
if a conservative
Jaccard similarity threshold is
    used to perform set containment search.\footnote{By conservative,
      we mean that we choose a Jaccard similarity threshold that
      guarantees zero false negatives.} 
   Our main contribution is
    presented in Section~\ref{sec:partition}.  We present a partitioning
    strategy that optimally minimizes the false positive 
rate
if an ideal 
Jaccard similarity
filter is used in each partition.
    We use dynamic LSH indexes to implement the
    Jaccard 
similarity
filter, as discussed in Section~\ref{sec:lsh-search}. We
    optimally configure the parameters so that
    the inherent false positive and 
false
negative errors associated with the LSH indexes
    are minimized.
}

Table~\ref{tbl:symbols} is a summary of the symbols used in the paper.

\subsection{Similarity Filtering for Containment}
\label{sec:conservative}

For the containment search problem, we are given a desired minimal
containment as a threshold $t^*$.  On the other hand, LSH can only be
tuned given a Jaccard similarity threshold $s^*$.  Thus, in order to use
LSH for containment search, the containment threshold needs to be
converted to a Jaccard similarity threshold.

Consider a domain $X$ with domain size $x=|X|$ and query
$Q$ with domain size $q=|Q|$.
We can convert Jaccard similarity and containment back and forth by \rjmnew{the}
inclusion-exclusion principle.
If $t=t(Q,X)$, we can compute the corresponding Jaccard similarity 
$s = s(Q,X)$ given the domain sizes $x$ and $q$.  
The transformations are given as follows.
\begin{equation}
    \label{eq:s-t}
    \Hs{x,q}(t) = \frac{t}{\frac{x}{q} + 1 - t},~
    \Ht{x,q}(s) = \frac{(\frac{x}{q}+1)s}{1+s}
\end{equation}

Notice the transformation 
$t^*\mapsto s^*$ depends on the domain size $x$, which is typically 
not a constant, 
so we need to  approximate $x$.  We choose to do so in a way
  that ensures the transformation to $s^*$ does not introduce any new false
  negatives over using $t^*$.
Suppose that $X$ is from a {\em partitioned} set of domains with
sizes in the interval of $[l, u)$, where
$l$ and $u$ are the lower and upper domain size bound of the partition.
A conservative approximation can be made by using the upper bound $u$ for $x$.
This ensures that filtering by $s^*$ will not result in any new false negatives.

We define a Jaccard similarity threshold using the upper bound $u$\eric{:}
\begin{equation}
\label{eq:jaccard-threshold}
    s^* = \Hs{u,q}(t^*) = \frac{t^*}{\frac{u}{q}+1-t^*}
\end{equation}
while the exact Jaccard similarity threshold is $\Hs{x,q}(t^*)$.
\eric{Because $u \ge x$ and $\Hs{x,q}(t^*)$ decreases monotonically with respect to 
$x$, we know $s^* = \Hs{u,q}(t^*) \le \Hs{x,q}(t^*)$.} 
Thus, by using this
approximation for $s^*$, we avoid false negatives introduced by the approximation.

The search procedure is described in
Algorithm~\ref{alg:search}. \rjmnew{The function} $\mathbf{approx}(|Q|)$ \rjmnew{provides an} estimation of the query domain size;
this can be done in constant time using the MinHash signature of 
the query~\cite{Cohen:PODC:2007}. We remark that the choice 
of $s^*$ will not yield any false negatives, provided
that {\tt Similarity-Search} perfectly filters away all domains with Jaccard
similarity less than $s^*$. Of course, this may not be the case, but our
conservative choice of $s^*$ using $u$ guarantees no new false negatives are
introduced. We will describe the index $\mathbf{I}(\mathcal{D})$ and the {\tt
Similarity-Search} shortly.

\begin{algorithm}[t]
    \DontPrintSemicolon
    \LinesNumbered
    Containment-Search($\mathbf{I}(\mathcal{D})$, $\sig(Q)$, $t^*$) \;
    \KwIn{$\mathbf{I}(\mathcal{D})$: an index of a collection of domains
        $\sig(Q)$: the MinHash signature of a query domain
          $t^*$: containment threshold}
    let $l\leftarrow \min\{|X|: X\in \mathcal{D}\},\; u=\max\{|X|: X\in \mathcal{D}\}$ \;
    let $q\leftarrow \mathbf{approx}(|Q|)$ \;
    let $s^* = \Hs{u,q}(t^*)$ \;
    let $\mathcal{D}_\mathrm{candidates}$ = {\tt Similarity-Search}($\mathbf{I}(\mathcal{D}), Q, s^*$)\;
    \KwRet $\mathcal{D}_\mathrm{candidates}$
    \caption{Domain Search with Set Containment}
    \label{alg:search}
\end{algorithm}

An LSH Ensemble partitions the set of {\em all} domains in $\mathcal{D}$ by their domain size into disjoint intervals.
Let $[l_i, u_i)$ be the lower and upper bounds of the $i$-th partition, and
$\mathcal{D}_i = \{X\in\mathcal{D}: l_i\leq |X|< u_i\}$, where $i=1\cdots n$.
With $n$ partitions, we can search each partition and take the
union of the individual query answers. Algorithm~\ref{alg:search} provides 
the search procedure for a {\em single} partition.
{\small
\begin{eqnarray*}
    && \makebox{\tt Partitioned-Containment-Search}
    (\{\mathbf{I}(\mathcal{D})\}, \sig(Q), t^*) \\
    &=&\bigcup_{i}
        \makebox{\tt Containment-Search} 
            (\mathbf{I}(\mathcal{D}_i),\sig(Q), t^*)
\end{eqnarray*}}
The {\tt Partitioned-Containment-Search} can be evaluated 
concurrently where {\tt Containment-Search} can be evaluated in parallel, and their
results are unioned.

\subsection{A Cost Model for Containment Search}
\label{sec:costmodel}

Let the time complexities of {\tt Containment-Search} and {\tt
Similarity-Search} be denoted by $T_\mathrm{containment}$ and
$T_\mathrm{similarity}$ respectively.

The complexity of Algorithm~\ref{alg:search} is given by:
\begin{equation}
T_\mathrm{containment} = T_\mathrm{similarity} 
+ \Theta(\makebox{correct result})+ \Theta(\NFP_{l,u})
\label{eq:T}
\end{equation}

The last two terms are the cost of processing the query result.
The value $\NFP_{l,u}$ is the number of false positives from 
\\ {\tt Similarity-Search} in the partition $[l,u)$.
In Equation~\ref{eq:T}, we only want to minimize the term $\Theta(\NFP_{l,u})$
to reduce the overall cost.

To minimize the time complexity of the parallel evaluation of {\tt
Partitioned-Containment-Search}, we wish to minimize the following cost function
by designing the partitioning intervals $[l_i, u_i)$:
\begin{equation}
    \mathrm{cost} = \max_{1\leq i\leq n}\NFP_{l_i, u_i}
    \label{eq:cost}
\end{equation}

\rmj{Before we can use this cost model, we must be able estimate the
  number of false positive in a partition.  This cost model and FP
  estimation will allow us to compute the expected cost of a particular
  partition and develop an optimal partitioning.
} 

\subsection{Estimation of False Positives}
\label{sec:estimatefp}
By using Jaccard similarity filter with threshold $s^*$ instead of the
containment threshold $t^*$, we incur false positives in the search result of
{\tt Similarity-Search} even if no error is introduced by using MinHash
signatures and $\mathbf{I}(\mathcal{D})$.

Consider a domain $X$ with size $x$ in some partition $[l, u)$.  
Domain $X$ would be a false positive if it meets the passing
condition of the approximated Jaccard similarity threshold, but actually fails
the containment threshold.

\begin{figure}
    \includegraphics[width=\columnwidth, natwidth=1, natheight=1]{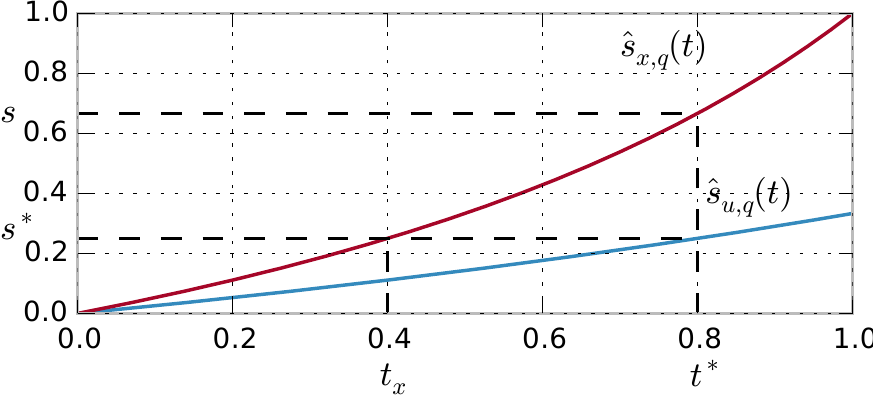}
    \caption{The relationships among $t_x$, $t^*$, and $s^*$}
    \label{fig:FP}
\end{figure}

Let the containment of the domain be $t$, and the Jaccard similarity be $s$.
The domain is a false positive if $t < t^*$ but $s > s^*$.
{\tt Similarity-Search} filters by $s^*$, so it is filtering $X$
according to an effective containment threshold given by
$\Ht{x,q}(s^*)$.  Define $t_x = \Ht{x,q}(s^*)$.
Thus, a domain is a false positive if its containment $t$ falls in the interval
of $[t_x, t^*]$.  Namely, its true containment score is below the query threshold
$t^*$, but above the effective threshold $t_x$.  The relationship between $t_x$
and $t^*$ is illustrated by Figure~\ref{fig:FP},
\eric{which plots the Jaccard similarities $\Hs{x,q}(t)$ and $\Hs{u,q}(t)$ as functions 
of containment $t$, with $u=3$, $x=1$ and $q=1$.}

\begin{prop}
\label{prop:effectivethreshold}
The effective containment threshold for $X$ is related to the query
containment threshold by the following relation.
\begin{equation}
t_x = \frac{(x+q) t^*}{u+q}
\end{equation}
\end{prop}

Given no prior knowledge about the domain $X$, we assume its containment
is uniformly distributed in the interval $[0, 1]$.
Thus, we can estimate the probability 
of
a true negative being falsely identified
as a candidate.
\begin{equation}
    P(X~is~\FP) = (t^* - t_x)/t^*
\end{equation}

Let $N_{l,u}$ be the expected total number of domains with sizes in the interval 
$[l, u)$. The expected number of false positives produced by the threshold is
given by
\begin{equation}
     \NFP_{l,u} = \sum \{P(X~is~\FP) : X\in\mathcal{D}, l\leq |X|< u\}
    \label{eq:FP1}
\end{equation}
where $N_{l,u}$ is the total number of domains that belong to the
partition $[l,u)$.

If we further assume that, within a partition, the distribution of $x$ is
uniform,
then we can evaluate Equation~\ref{eq:FP1} further to obtain the
following result.
\begin{prop}
    Assuming a uniform distribution of domain sizes in the interval $[l, u)$,
    the upper bound of the number of candidate domains which
    are false positives is given by
    \begin{equation}
        \NFP_{l,u} \leq N_{l,u}\cdot \frac{u-l+1}{2u} 
    \label{eq:FP}
    \end{equation}
    \label{prop:2}
\end{prop}
\proof (Outline)
We need to cover several cases: (1) $ t^*q \leq l$, (2) $t_lq \leq l < t^*q$ and $ t^*q \leq u$, (3) $l < t_lq$ and $ t^*q \leq u $, (4) $l < t_l q$ and $t_uq \leq u < t^* q$, and finally (5) $u < t_u q$.  For the first case $t^*q \leq l$, we have the probability of a domain with size $x$ being a false positive being $(t^*-t_x)/t$.
\begin{align}
	N_{l,u}^\mathrm{FP} &= N_{l,u}\sum_{x=l}^{u-1} \frac{t^*-tx}{t^*}\frac{1}{u-l} dx\\
	&= N_{l,u}\cdot\frac{u-l+1}{2(u+q)} 
    \leq N_{l,u}\cdot \frac{u-l+1}{2u}   
\end{align}
If $u \gg q$, then the upper bound is a tight upper bound.
The other cases (2-5) are proven similarly.
\qed\\

\eric{To summarize, by assuming uniform distribution of domain size
within an interval, we can compute an upper bound of the number of false positives
in the interval given only 
its boundaries $[l, u)$ and its number of domains $N_{l,u}$.}

\subsection{Optimal Partitioning}
\label{sec:partition}

The cost function defined in Equation~\ref{eq:cost} and the estimate of false
positive candidate domains in Equation~\ref{eq:FP} allow us to compute the
expected cost of a particular partition.  In this section, we
provide a concrete construction of a minimal cost partitioning.

We denote a partitioning of the domains in $\mathcal{D}$ by $\Pi =
\left<[l_i, u_i)\right>_{i=1}^n$ where $l_{i+1} = u_i$.   
Let $\NFP_i = \NFP_{l_i, u_i}$. An optimal partitioning is then one that
minimizes the number of false positives over all possible
partitions.
\begin{definition}[The Optimal Partitioning Problem]
   A partitioning $\Pi^*$ is an optimal
    partitioning if $$\Pi^* = \argmin_{\Pi}\max_{i}\NFP_i$$
\end{definition}

We now show that there is an optimal partitioning that has
  equal number of false positives in each partition.
\begin{theorem}[Equi-FP optimal partitioning]
    There exists an optimal partitioning $\Pi^*$ such that
    $\NFP_i = \NFP_j$ for all $i, j$.
    \label{thm:1}
\end{theorem}
\proof (Outline)
Recall that the cost function is
$\cost(\Pi) = \max_i\NFP_i$.
We can show that $\NFP_i$ is monotonic with respect to $[l_i, u_i)$.
For simplicity, suppose we only have two partitions, $n=2$.  Suppose we
have an optimal partitioning $\Pi_1$ such that $\NFP_1 \neq \NFP_2$.  
Without loss of generality, we assume that $\NFP_1 < \NFP_2$.  By the
monotonicity of $\NFP$ w.r.t. the interval, we can increase the value
of $u_1$ such that $\NFP_1$ is increased and $\NFP_2$ is decreased until
$\NFP_1 = \NFP_2$.
The new partitioning $\Pi_2$ obtained by the adjustment of $u_1$ is such that
$\cost(\Pi_2)\leq \cost(\Pi_1)$, so it is also optimal.
This proof generalizes to arbitrary $n$.  \qed \\

As a result of Theorem~\ref{thm:1}, we can construct an equi-$\NFP_i$ partitioning
$\Pi$ as a way to optimally partition the domains.  Unfortunately, this means
that $\Pi$ is query dependent because $\NFP_i = \NFP_i(q)$.  We cannot afford
to repartition $\mathcal{D}$ for each query.
Fortunately, with a reasonable assumption, there exists a {\em query independent} way
of partitioning $\mathcal{D}$ near-optimally.  We assume that $\max\{|X|:
X\in\mathcal{D}\} \gg q$.  Namely, $\mathcal{D}$ contains {\em large} domains
relative to the query.  
An index will most
often be used with queries that are much smaller than the maximum domain size.

Let $M_i$ be the upper bound on $\NFP_i$ following Proposition~\ref{prop:2}:
\begin{equation}
M_i = N_{l_i,u_i}\cdot\frac{u_i-l_i+1}{2 u_i}
\label{eq:NFP-UB}
\end{equation}

Let $\Pi^*$ be an optimal equi-$\NFP_i$ partitioning whose existence is guaranteed
by Theorem~\ref{thm:1}.
Under the large domain assumption, we can see that 
\begin{equation}
\cost(\Pi^*) = \NFP_n \approx M_n
\label{eq:cost-Mn}
\end{equation}
which is query independent. This suggests that we can {\em approximate} the optimal
partitioning $\Pi^*$ by an equi-$M_i$ partitioning of $\mathcal{D}$.
\eric{
Furthermore, we show that for power-law distribution, equi-$M_i$ partitioning is the
same as equi-$N_{l_i,u_i}$, or {\it equi-depth}.
}

\eric{
\begin{theorem}[Partitioning and power-law]
    Suppose $\mathcal{S}=\{|X|: X\in \mathcal{D}\}$ 
    follows a power-law distribution with
    frequency function $f(x) = Cx^{-\alpha}$ where $\alpha > 1$.
    The equi-$N_{l_i,u_i}$ partitioning
    \begin{equation}
        \Pi = \left<[l_i, u_i)\right>_{i=1}^n,~N_{l_i,u_i}=|\mathcal{S}|/n,~\forall i=1\dots n
    \end{equation}
    is an equi-$M_i$ partitioning.
    \label{thm:2}
\end{theorem}
\proof 
We use the (tight) bound on number of false positives defined by Equation~\ref{eq:NFP-UB}.
Using the property of equi-$N_{l_i,u_i}$ partitioning, we have
\begin{equation}
    N_{l_i,u_i} = \int_{l_i}^{u_i}Cx^{-\alpha}dx = |\mathcal{S}|/n,~\forall i=1\dots n
\end{equation}
Solving the equation above we obtain a relationship between $l_i$ and $u_i$: 
\begin{equation}
    \frac{l_{i}}{u_{i}}= \left|1-\frac{|\mathcal{S}|(\alpha-1)l_{i}^{\alpha-1}}{nC}\right|^{-(\alpha-1)}=g(l_{i}), ~ \forall i=1\dots n
\end{equation}
With the expressions for $u_i$ and $l_i$, we get
\begin{equation*}
    \frac{u_i - l_i+1}{2u_i} = \frac{1}{2}-\frac{g(l_i)}{2} + \frac{1}{2u_i} 
    \approx \frac{1}{2}
\end{equation*}
Since $g(l_i)$ and $\frac{1}{2u_i}$ approach 0 fast 
with increasing $l_i$ and $u_i$ respectively,
and are likely very small comparing to $\frac{1}{2}$ in practice 
when $|S|$ is in the millions and $2u_i \gg 1$, 
we can discard them, and obtain a constant for all partitions.
Given that both terms in Equation~\ref{eq:NFP-UB} are constants,
the equi-$N_{l_i,u_i}$ partitioning is an equi-$M_i$ partitioning.
\qed\\}

\eric{
Theorem~\ref{thm:2} provides a fast method for approximating the optimal partitioning.
For a collection of domains, 
we can create a desired number of partitions such that all contain the
same number of domains. 
Based on Theorem~\ref{thm:2}, if the domain size distribution
follows a power-law, this is an equi-$M_i$ partitioning
and hence, a good approximation to equi-$\NFP_i$ partitioning, which is optimal.}

\subsection{Containment Search Using Dynamic LSH}
\label{sec:lsh-search}

In Algorithm~\ref{alg:search}, we assume the existence of a search algorithm
based on a Jaccard similarity threshold for {\em each} partition $\mathcal{D}_i$.
We choose to index the domains in $\mathcal{D}_i$ using a MinHash LSH index
with parameters $(b, r)$ where $b$ is the number of bands used by the LSH index,
and $r$ is the number of hash values in each band.

Traditionally, MinHash LSH has a fixed $(b, r)$ configuration, and thus has a
static Jaccard similarity threshold given by the approximation:
\begin{equation}
s^* \approx (1/b)^{(1/r)}
\label{eq:jthreshold-br}
\end{equation}
One
can observe that MinHash LSH only approximates \\{\tt Similarity-Search}, and
therefore will introduce false positives, in addition to $\NFP$ in each
partition, and also false negatives.  In this section, we describe our method
of selecting $(b, r)$ so that the additional error due to the approximation
nature of MinHash LSH is minimized.
Since $s^*$ is query dependent, we choose to use a dynamic LSH index, such as
the LSH Forest~\cite{Bawa:WWW:2005}, so we can vary $(b, r)$
for each query.
LSH Forest uses a prefix tree to store the $r$ hash values in each band (see 
Section~\ref{sec:lsh}), thus the effective value of $r$ can be changed at query
time by choosing the maximum depth to traverse 
to
in each prefix tree.
The parameter
$b$ can be varied by simply choosing the number of prefix trees to visit.

\begin{figure}
    \includegraphics[width=\columnwidth,natwidth=1, natheight=1]{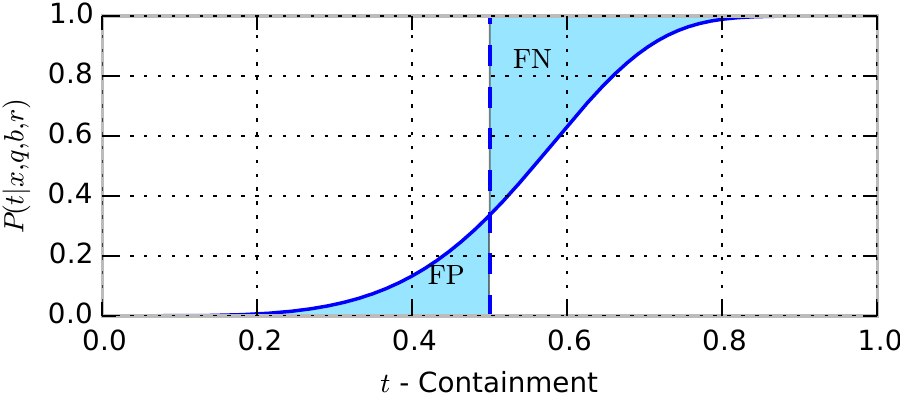}
    \vspace{-1mm}
    \caption{$P(t|x,q,b,r)$ - probability of becoming candidate with
	respect to containment, given $x=10$, $q=5$, $b=256$, and $r=4$;
containment threshold $t^*=0.5$ (dashed line)}
    \label{fig:lshtcurve}
\end{figure}

We now
describe how the parameters $(b, r)$ are selected so that
the resulting selectivity agrees with $s^*$.
Using the relationship between Jaccard similarity $s$ and containment $t$,
we express the probability of a domain $X$ becoming a candidate in terms of
$t=t(Q,X)$, $x=|X|$, and $q=|Q|$ instead of $s=s(Q,X)$.
\begin{equation}
    \label{eq:probLSH}
    P(t|x,q,b,r) = 1-\left(1-s^r\right)^b = 1 -\left(1 - \left(\frac{t}{\frac{x}{q}+1-t}\right)^r\right)^b
\end{equation}
Figure \ref{fig:lshtcurve} plots the probability 
(for $t^*=0.5$), along with the areas
corresponding to the false positive ($\FP$) and false negative probabilities ($\FN$)
induced by the MinHash LSH approximation. 
It is important to notice that the 
false positives here are introduced by the use
of {\tt Similarity-Search} with MinHash LSH, which is
different from the false positive introduced by the transformation of the containment
threshold to a Jaccard similarity threshold in Section~\ref{sec:estimatefp}. 

Note that $t$ cannot exceed the size ratio $x/q$. 
We can express the probability of $X$ being a false positive, $\FP$, or a false
negative, $\FN$, in terms of $t^*$ and $x/q$.
\begin{align}
    \label{eq:probFP}
    \FP(x,q,t^*,b,r) = \left\{\begin{matrix}
\int_{0}^{t^*} P(t|x,q,b,r)dt & \frac{x}{q}\ge t^* \\
\int_{0}^{\frac{x}{q}} P(t|x,q,b,r)dt & \frac{x}{q} < t^*
\end{matrix}\right.
\end{align}
\begin{align}
    \label{eq:probFN}
    \FN(x,q,t^*,b,r) = \left\{\begin{matrix}
\int_{t^*}^{1} 1-P(t|x,q,b,r)dt & \frac{x}{q} \ge 1 \\
\int_{t^*}^{\frac{x}{q}}1-P(t|x,q,b,r)dt & t^*\le\frac{x}{q} < 1\\
0  & \frac{x}{q} < t^*
\end{matrix}\right.
\end{align}
The optimization objective function for tuning LSH is given as:
\begin{equation}
    \label{eq:lshtuning}
    \argmin_{b,r} (\FN + \FP)(x,q,t^*,b,r), 
    \makebox{such that } 0 < b r \le m
\end{equation}
where $m$ is the number of minwise hash functions.
Since $x$ is not constant within a partition, we cannot use this objective
function for tuning the LSH of the partition.
As described in Section~\ref{sec:conservative}, for each partition $i$,
we used $u_i$ to approximate $x \in [l_i,u_i]$, and the alternative objective
function for tuning we used is:
\begin{equation}
    \argmin_{b,r} (\FN + \FP)(u_i,q,t^*,b,r), 
    \makebox{such that } 0 < b r \le m
\label{obj-func}
\end{equation}
For some particular value of $x$,
the $b$ and $r$ found using the alternative objective function
would be more optimal if $u_i$ were closer to $x$.
Formally, for some $\epsilon> 0$, there exists $\delta > 0$ such that when
$u_i - x < \delta$,
\begin{equation*}
    (\FP+\FN)(x,q,t^*,b_p,r_p) - (\FP+\FN)(x,q,t^*,b_\mathrm{opt},r_\mathrm{opt}) < \epsilon
\end{equation*}
where $b_p$ and $r_p$ are the parameters found using the alternative objective
function,
and $b_\mathrm{opt}$ and $r_\mathrm{opt}$ are the optimal parameters computed with the
exact objective function.

The computation of $(b,r)$ can be handled offline. Namely, we choose to
pre-compute the $\FP$ and $\FN$ for different combinations of $(b,
r)$. At query time, 
the
pre-computed $\FP$ and $\FN$ are used to 
optimize objective function in Equation~\ref{obj-func}.
Given that $b$ and $r$ are positive integers and their product must be less than
$m$, the computed values require minimal memory overhead.

Finally, we query each partition with the dynamically determined $(b, r)$ using a dynamic LSH index as described by Bawa et al.~\cite{Bawa:WWW:2005}.

In summary, the query evaluation of LSH Ensemble performs dynamic transformation of the containment threshold to a per-partition threshold of Jaccard similarity.  Then we query each partition with different $(b, r)$ to identify the candidate domains.

 \section{Experimental Analysis}
\label{sec:experiment}

We evaluate the accuracy and performance of our LSH Ensemble technique,
and compare against the \kenpu{traditional MinHash LSH~\cite{Indyk:STOC:1998} and
the state-of-art Asymmetric Minwise Hashing~\cite{Shrivastava:WWW:2015} as
baselines}.

\kenpu{
    We have collected the relational data from the Canadian Open Data Repository
    (as of June 2015) which consists of 10,635 relations with 65,533 domains.
    The size of this data set allows us to compute the ground truth for any
    domain search queries, thus allowing us to reliably measure the accuracy of
    LSH Ensemble and 
the two baselines
in Section~\ref{sec:accuracy}.}
\rmj{We evaluate the effectiveness of LSH Ensemble over dynamic data
  by showing the index is robust to changes in the distribution of the
  domain sizes (Section~\ref{sec:dynamic}).}

\kenpu{
    To test the performance LSH Ensemble at Internet scale, we used the entire
    English portion 
    of the WDC Web Table Corpus 2015~\cite{Lehmberg:WWW:2016} which 
    consists of 51 million relations with 262 million domains (Section~\ref{sec:scale}).
    The LSH ensemble for the English Web Table corpus span over a cluster with
    5 nodes, each with 8 Xeon processors and 64 GB memory.}

We list the variables used in our experiments in Table~\ref{tbl:variables},
with default values in bold face.

\begin{table}[]
        \centering
        \caption{Experimental Variables}
        \label{tbl:variables}
        \begin{tabularx}{\columnwidth}{@{} l Y c @{}}
        \toprule
        Variable & Range \\
        \midrule
         Num.~of Hash Functions in MinHash ($m$) & {\bf 256} \\ \addlinespace
        Containment Threshold ($t^*$) & 0 - {\bf 0.5} - 1 \\ \addlinespace
        Num.~of Domains $|\mathcal{D}|$ & 65,533 - 262,893,406  \\ \addlinespace
        Num.~of Queries & {\bf 3,000} \\ \addlinespace
        Num.~of Partitions ($n$) & 8-{\bf 32} \\ \addlinespace
        Skewness & 0.50 - {\bf 13.87} \\
        \bottomrule
        \end{tabularx}
\end{table}
\vspace{-2mm}
\subsection{Accuracy of LSH Ensemble}
\label{sec:accuracy}

\begin{figure*}[t]
        \includegraphics[width=\textwidth, natwidth=1, natheight=1]{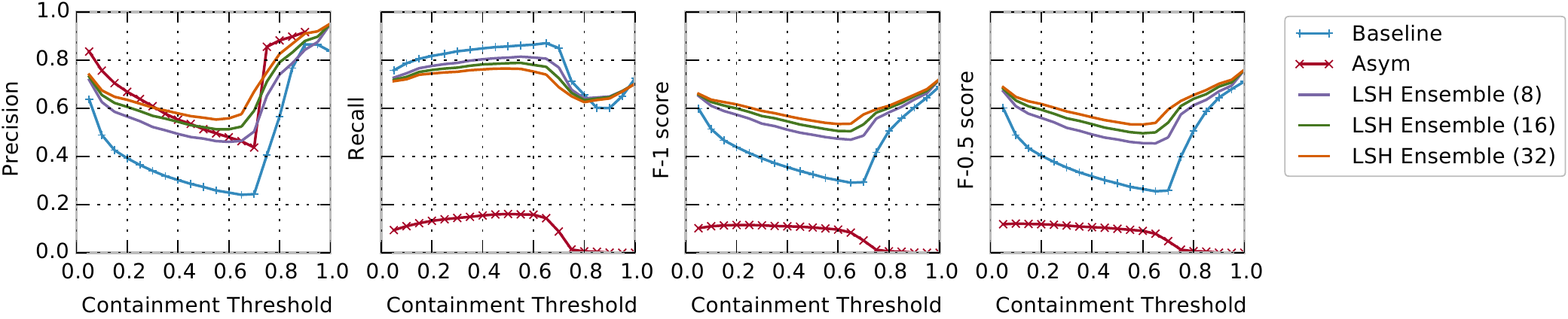}
\vspace{-1mm}
        \caption{Accuracy versus Containment Threshold on Canadian Open Data Corpus}
        \label{fig:pr}
\end{figure*}
\begin{figure*}[t]
        \includegraphics[width=\textwidth, natwidth=1, natheight=1]{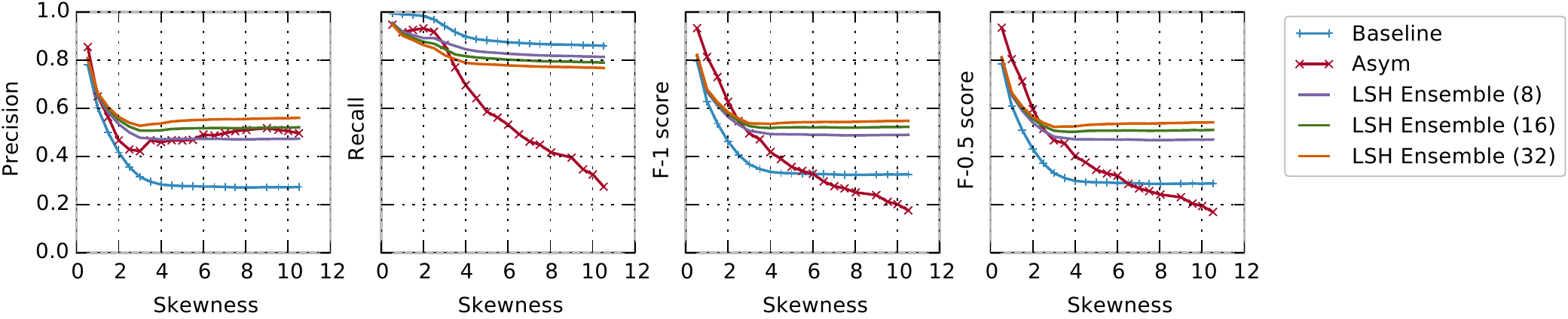}
\vspace{-1mm}
        \caption{Accuracy versus Domain Size Skewness}
        \label{fig:pr_skew}
\vspace{-1mm}
\end{figure*}

For our experimental evaluation,
we use the set-overlap based definition of precision and recall.
Let $\mathcal{D}$ be the set of domains in the index.
Given a query domain $Q$ and a containment threshold $t^*$,
the ground truth set is defined as
$T_{Q, t, \mathcal{D}} = \{ X | t(Q, X) \ge t^*, X \in \mathcal{D} \}$.
Let $A_{Q, t^*, \mathcal{D}}$ be the set of domains returned by \eat{the}\rjmnew{a} search algorithm.
Precision and recall are defined as follows.
\begin{equation}
	Precis. = \frac{|A_{Q,t^*,\mathcal{D}} \cap T_{Q,t^*,\mathcal{D}}|}{|A_{Q,t^*,\mathcal{D}}|},
	Recall = \frac{|A_{Q,t^*,\mathcal{D}} \cap T_{Q,t^*,\mathcal{D}}|}{|T_{Q,t^*,\mathcal{D}}|}
\end{equation}
\eric{We also used $\mathrm{F}_{\beta}$ score to evaluate the overall accuracy. We set $\beta$ to 1 (equality weighted)
and 0.5 (precision-biased). Since our algorithm is recall-biased, assigning more weight to precision gives a fairer evaluation of the overall accuracy.}
\begin{equation}
	\mathrm{F}_{\beta} = \frac{(1 + \beta^{2}) \cdot Precis. \cdot Recall}{ \beta^{2} \cdot Precis. + Recall}
\end{equation}

To evaluate accuracy,
we \eat{run}\rjmnew{use}\
three main experiments over
the CSV files from the Canadian Open Data repository \rjmnew{for which we computed ground truth (exact containment scores)}.
We did not use the WDC Web Table Corpus
due to the extremely high time cost of obtaining the ground truth.
We discarded domains with fewer than ten values and obtained 65,533
such domains which we indexed.
Then, we sampled a subset of 3,000 domains and used them as queries.
As we noted in Section~\ref{sec:intro},
the domain size distribution of the Canadian Open Data roughly follows a
power-law distribution (shown in Figure~\ref{fig:card_dist}).
Thus, we applied the equi-depth partitioning based on Theorem~\ref{thm:2}.

\rjmnew{For a fair comparison,} all the indexes including MinHash LSH and Asymmetric Minwise Hahsing \rjmnew{are implemented to use} the dynamic LSH \eat{implementation}\rjmnew{algorithm} for containment
search
described in Section~\ref{sec:lsh-search}, and the upper bound
of domain sizes is used to convert containment threshold to Jaccard similarity
threshold as described in Section~\ref{sec:conservative}.

\begin{figure*}[t]
        \includegraphics[width=\textwidth, natwidth=1, natheight=1]{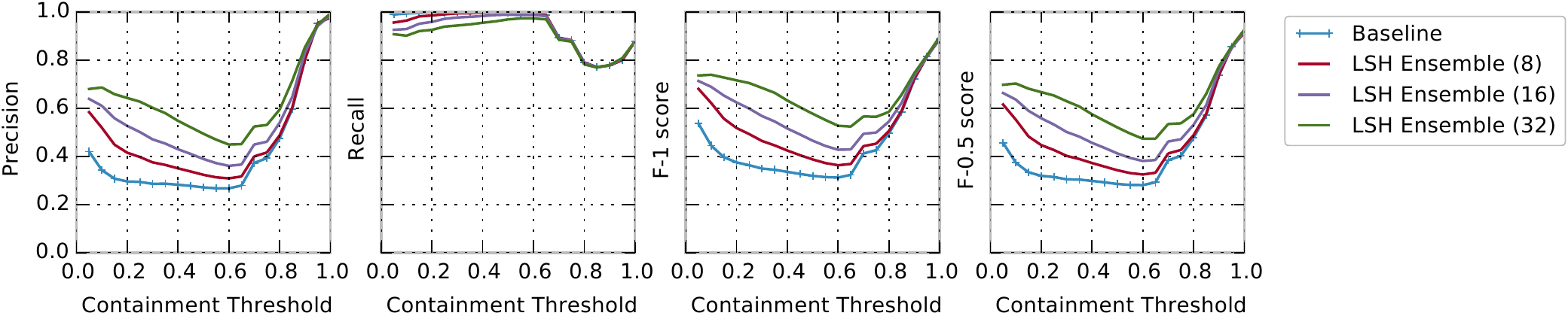}
\vspace{-2mm}
        \caption{Accuracy of Queries with Large Domain Size (Highest 10\%)}
        \label{fig:pr_large}
\end{figure*}
\begin{figure*}[t]
        \includegraphics[width=\textwidth, natwidth=1, natheight=1]{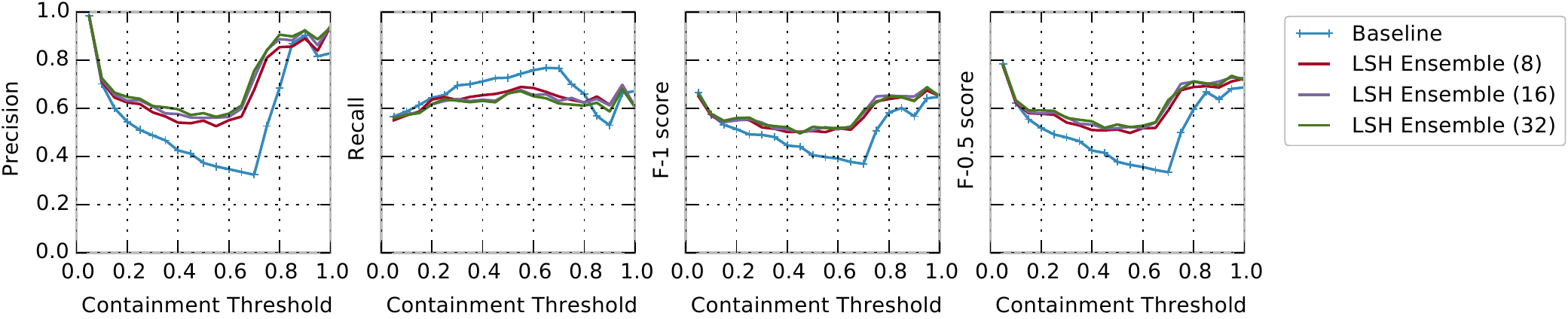}
\vspace{-2mm}
        \caption{Accuracy of Queries with Small Domain Size (Smallest 10\%)}
        \label{fig:pr_small}
\end{figure*}
\vspace{-2mm}
\vspace{4mm}
\noindent
\textbf{Impact of Partitioning.}
Figure~\ref{fig:pr} shows the precision, recall, F-score and F$_{0.5}$-score of
MinHash LSH (Baseline), Asymmetric Minwise Hashing (Asym), and
LSH Ensembles with different numbers of partitions
(i.e., 8, 16, and 32).
We report the average precision and recall for
every containment threshold from 0.05 to 1.0 with a step size of 0.05.
The equi-depth partitioning provides a clear improvement in precision
over the baseline, verifying our
theoretical
analysis \eat{on}\rjmnew{of} partitioning.
As the number of partitions grows,
the precision increases for all containment thresholds;
however the relative gain diminishes as the upper bound of each
partition becomes a better approximation of the domain sizes of the partition.
Recall decreases by about 0.02 each time the number of partitions doubles.
\eric{The false negatives are introduced by the MinHash LSH in each partition,
as described in Section~\ref{sec:lsh-search}.}
\eric{The index tuning} becomes less conservative
(regarding false negatives) as the upper bound of a partition
approaches the actual domain sizes.
\eric{There is a trade-off between partitioning and recall, however,}
since the approximation always remains conservative, the recall is not
affected significantly.
\eric{The $\mathrm{F}_{\beta}$ score results show that LSH Ensemble improved the overall accuracy
over the baseline by as much as 25\%.}
\eric{Since the threshold changes the optimizaiton constraints for index tuning
(see Section~\ref{sec:lsh-search}) and the parameter space is integer,
discontinuity may exist in the optimization objective. This leads to
the sudden increase in precision at threshold 0.7.}

Asymmetric Minwise Hashing achieved high precision comparable to that of LSH Ensemble, however,
performed poorly in recall.
Due to padding, some of the hash values in a MinHash signature are from the new padded values.
For an indexed domain to become a candidate, its hash values from the original domain need
to have collisions with the hash values in the signature of the query domain.
Thus, \rjmnew{with a finite number of hash functions}, the probability of an indexed domain becoming a candidate,
including \eat{the ones}\rjmnew{domains} that should be qualified, \eat{are diminished}\rjmnew{is lowered}.
This issue is not significant when the skew in domain sizes is small.
However, when the skew is high, the amount of padding required becomes very large, making the probability
of qualifying domains becoming candidates very low and sometimes nearly zero.
This explains the low average recall 
over the skewed Open Data domains.\footnote{\small{\eric{See the technical report for detailed analysis of Asymmetric Minwise Hashing: {\tt http://arxiv.org/abs/1603.07410}}}}
As the containment threshold increases, domains need to have a greater number of collisions in their hash values.
Thus, the probability of finding qualifying domains drops to zero for high thresholds, resulting in a recall of zero.
\eric{Around 80\% of Asymmetric Minwise Hashing query results are empty for thresholds up to 0.7,
more than 98\% are empty for thresholds higher than 0.8, and 100\% are empty for thresholds 0.95 and 1.0.
We consider an empty result having precision equal to 1.0, however, we exclude such results when computing average precisions.}
For domain search, a high recall is a necessity.
Thus, \eat{we suggest}\rjmnew{these results show} that LSH Ensemble is a better choice for the domain search problem \rjmnew{over skewed data}.

\vspace{1mm}
\noindent
\textbf{Impact of Query Size.}
The equi-depth partitioning assumes the domain size of the query is much
smaller than the maximum domain size of the \rjmnew{partitioned} index.
We investigated whether in practice large query domain size influence the
effectiveness of the partitioning technique.
Figure~\ref{fig:pr_small}
shows the precision and recall \eat{of the}\rjmnew{for} queries with domain sizes in the smallest 10\%,
and Figure~\ref{fig:pr_large}
shows the precision and recall \eat{of the}\rjmnew{for} queries with domain sizes in the largest 10\%.
For the large queries, the precision is smaller, due to the assumption no longer holding.
Still, the precision increases with more partitions, confirming our analysis that
partitioning should always increase precision, and the recall stays high.
The result of the small queries is similar to the overall result in Figure~\ref{fig:pr},
this is likely due to the fact that the domain sizes of the queries
also follows a power-law distribution, so the queries contained mostly small domains.

\vspace{1mm}
\noindent
\textbf{Effect of Skewness on Accuracy.}
In order to investigate the effect of skew on accuracy.
We created 20 subsets \rjmnew{of the Canadian Open Data}.
\rjmnew{The first contained a small (contiguous) interval of domain sizes.  We then expanded the interval repeatedly to create 19 larger subsets.}
We measured the precision
and recall of the MinHash LSH (Baseline), Asymmetric Minwise Hashing (Asym),
and LSH Ensemble on each
subset.
Since the domain sizes approximately follow a power-law,
the skewness increases as we use subsets with larger \eat{domain size interval}\rjmnew{intervals of domain sizes}.
\rjmnew{We compute the} skewness \rjmnew{of each subset using}
the following equation.
\begin{equation}
    \label{eq:skewness}
    skewness = \frac{m_{3}}{m_{2}^{3/2}}
\end{equation}
where $m_{2}$ and $m_{3}$ are the 2nd and 3rd moments \rjmnew{of the distribution}~\cite{Kokoska:2000}.
A higher positive skewness means there is more weight in the left tail of the distribution.

Figure~\ref{fig:pr_skew} shows that as skewness increases, the precision of all indexes
decreases, while a high level of recall is maintained except for Asymmetric Minwise Hashing.
This is expected for MinHash LSH. Because it uses the upper bound of domain
sizes in the conversion of containment to Jaccard similarity threshold
(see Section~\ref{sec:conservative}),
as the upper bound of domain sizes increases, the approximation becomes less precise,
increasing the false positive rate.
The same reason for the decrease in precision applies to LSH Ensemble as well,
however, the issue is less severe.
This is because of partitioning - the upper bound of each partition is a much better approximation
than the overall upper bound.
We can also observe that as the number of partitions goes up, the index is less affected
by skew.
\rjmnew{As expected,}
Asymmetric Minwise Hashing achieved high recall when the skewness is small,
but the recall decreases significantly as skewness increases.
This is again due to the large amount of padding when the skewness is high.

\kennew{We have also conducted experiments on evaluating the performance of
using Asymmetric Minwise Hashing in conjunction with partitioning (and up to 32 partitions).
Namely, we used Asymmetric Minwise Hashing instead of MinHash LSH in each partition.
While there is a slight improvement in precision, we failed to observe any significant improvements in recall.
This is due to the fact that, for a power-law distribution, some partitions
still have sufficiently large difference between the largest and the smallest domain sizes,
making Asymmetric Minwise Hashing unsuitable.
}

\subsection{\rmj{Dynamic Data}}
\label{sec:dynamic}
\rmj{Web data will certainly be dynamic.  While our index can easily
  accommodate the addition of new domains, if the new domains follow a
  different domain size distribution, our partitioning strategy may
  become less optimal (in particular, the partitions may start to have
  different sizes).  
Using domains from the Canadian Open Data, we simulated this scenario by creating partitionings 
that increasingly deviate away from equi-depth and move toward
equi-width (meaning all partitions have the same interval size).
We measured precision and recall of each partitioning, 
and computed the standard deviation among partition sizes.}

\begin{figure}[t]
        \includegraphics[width=\columnwidth, natwidth=1, natheight=1]{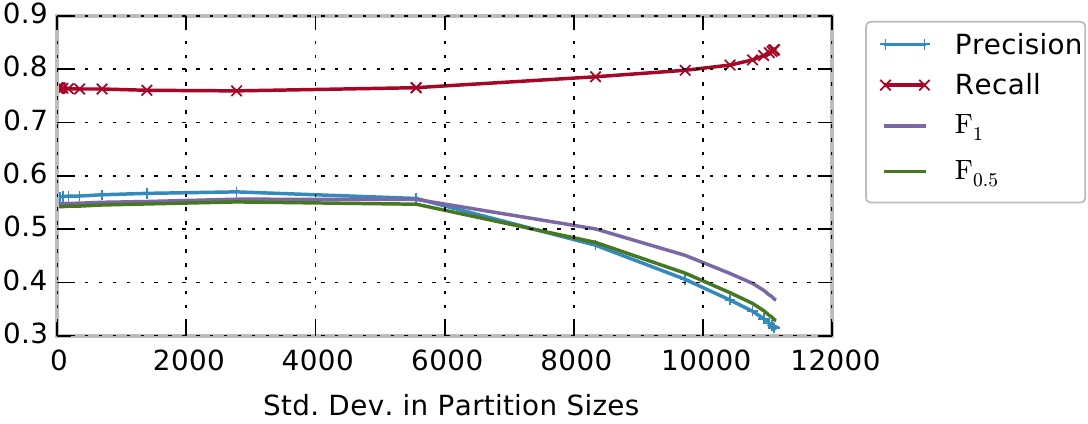}
        \caption{\eric{Accuracy vs. Std. Dev. of Partition Sizes}}
        \label{fig:robustness}
\end{figure}

\rmj{Figure~\ref{fig:robustness} shows that
as the partitioning deviates away from equi-depth, the standard
deviation increases.
However, the precision remains almost the same until the standard
deviation is increased beyond 5,556.  This is more than 2.7 times the
equi-depth partition size (of 2,047).
This represents a huge shift in the domain size distributions before
any degradation in accuracy is seen.  
So while the index may need to be rebuilt if the domain size
distribution changes drastically, this would be a rare occurrence.}

\subsection{Efficiency of Indexing and Searching}
\label{sec:scale}
For the performance experiments, we used the English relational subset of the
WDC Web Table Corpus 2015, from which we extracted 262,893,406 domains.
The domain size distribution closely resembles a power-law distribution, as shown
in Figure~\ref{fig:card_dist}.
Thus, we used the equi-depth partitioning for LSH Ensemble as suggested by
Theorem~\ref{thm:2}.
We selected a random subset of 3,000 domains to use as test queries.
\eric{Due to the massive number of domains, an index for all domains does not
fit in the memory of a single machine. 
Thus, we divided the domains into 5 equal chunks
on 5 machines, 
and then built an index for each chunk.
A query client sends query requests to all indexes in parallel, and the results are unioned.}

\rmj{Table~\ref{tbl:scalability} compares the indexing and query cost of the basic
MinHash LSH (baseline) with LSH Ensemble with different numbers of partitions.
The indexing cost of all indexes are very similar.
Since all partitions are indexed in parallel, the time 
to index
does not increase with the number of partitions.
On the other hand,
the query cost of LSH Ensemble is significantly smaller than the MinHash LSH.
\rjmnew{In part of course, this is due to the parallelism, but in addition, we are seeing speed up because the partitioning improves precision.}
The index becomes more selective as the number of partitions increases.  \rjmnew{For 8 partitions, most of the speedup is parallelism, but for 16 partitions, the improved precision also improves the performance dramatically.}
Since all partitions are queried in parallel, the number of partitions
does not contribute to query cost.}

\begin{figure}[t]
	\includegraphics[width=\columnwidth,natwidth=1, natheight=1]{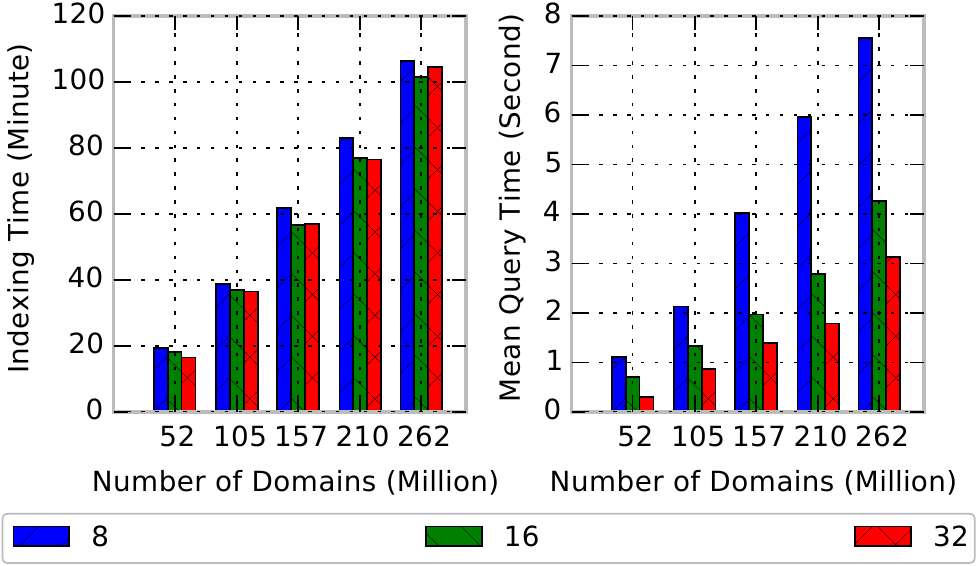}
	\caption{\eric{Indexing and Mean Query Cost}}
	\label{fig:scalability}
\end{figure}

\rmj{The first plot in Figure~\ref{fig:scalability} shows the indexing cost
with respect to the number of domains indexed.
The indexing performance of LSH Ensemble scales linearly with respect
to the number of domains.
In addition, because of the parallel implementation, the number of partitions
does not affect the indexing cost.
The second plot in Figure~\ref{fig:scalability} shows the query cost
of LSH Ensemble.
The query cost of LSH Ensemble increases with the number of domains in the index,
because the number of candidates returned also increases (for a given threshold),
and query processing cost includes the cost of outputting the candidates.
On the other hand, the query cost increases much slower with more partitions.
Again, this is because the precision is improved by partitioning, yielding a smaller selectivity.}

\begin{table}[t]
	\begin{tabular}{c | c | c}
		& Indexing (min)  & Mean Query (sec) \\ [0.5ex]
		\hline
		Baseline          & 108.47 & 45.13 \\
		LSH Ensemble (8)  & 106.27 & 7.55 \\
		LSH Ensemble (16) & 101.56 & 4.26 \\
		LSH Ensemble (32) & 104.62 & 3.12 \\
	\end{tabular}
	\vspace{-1mm}
	\caption{\eric{Indexing and Query Cost of
			Baseline and LSH Ensemble on 263 Milion Domains}}
		\label{tbl:scalability}
\end{table}

 \section{Conclusion and Future Work}
\label{sec:conclusion}

\kenpu{We proposed the {\em domain search problem} 
where the goal is to find domains that maximally contain a query domain
from Open Data repositories and other structured sources on the Web.}
We presented {\em LSH Ensemble}, a new index structure based on MinHash LSH, 
as a solution to \rjmnew{the} domain search problem using set
containment.
By means of partitioning, we \rjmnew{show how we} can
efficiently perform set containment-based queries even on hundreds of millions
of domains
with highly skewed distribution of domain sizes.  We constructed a cost model
that describes the precision of LSH Ensemble with a given partition.  \eat{It
was shown}\rjmnew{We show} that for any data distribution, there exists an
optimal partitioning scheme that equalizes the false positives across the
partitions. Furthermore, for datasets with a power-law distribution, the optimal
partitioning \eric{can be approximated using equi-depth}, amenable to efficient
implementation.

We have conducted extensive evaluation of LSH Ensemble on the Canadian Open Data
repository and the \kenpu{entire English} \eric{relational} WDC Web Table Corpus.  Our solution is
able to sustain query
times of \kenpu{few seconds} even at 262 million domains.  Compared to
other alternatives, LSH Ensemble consistently \rjmnew{performs
better} in both accuracy and performance.  In some cases, LSH Ensemble is
$\sim15$ times faster than a MinHash LSH index for set containment queries.

\vspace*{-0.75mm}
\section{Acknowledgment}
This work was partially supported by NSERC. 
\bibliographystyle{abbrv}
\bibliography{main}
\balancecolumns

\section*{Appendix}
\subsection*{Proof of Proposition 2}

Recall from Section~\ref{sec:estimatefp},
$N_{l,u}$ is the total number of domains in the
interval $[l, u)$.
The expected number of false positives in the interval produced by using the effective 
containment thresholds is given by:
\begin{align*}
	\NFP_{l,u} &= \sum \{P(X~is~\FP) : X\in\mathcal{D}, l\leq |X|< u\}\\
		   &= N_{l,u} \sum_{x=l}^{u-1} P(\FP|x) P(x) 
\end{align*}
where $P(\FP|x)$ is the probability of a domain becoming a false positive given its 
size is $x$,
and $P(x)$ is the probability density function of the domain size distribution
in the interval.
The effective containment threshold for a domain with size $x$ is given in 
Proposition 1 as:
\begin{equation*}
	t_x = \frac{(x+q) t^*}{u+q}
\end{equation*}
where $q$ is the query domain size, and $t^{*}$ is the query-given containment 
threshold.

For any domain with size $x$, its containment score is bounded by 
$\min \left\{ 1.0, x/q \right\}$.
Thus, given different $q$, the maximum and minimum containment scores from domains
in the interval can be different, and the probability of false positive can be
different as well.
We separate the proof into 5 cases.

{\bf Case 1: $t^{*}q \leq l$}.
Since the lower bound of the interval is greater than minimum domain size required to satisfy 
the containment threshold, all domains in the interval have non-zero probabilities of 
becoming false positives.
The false positive probability is given as
\begin{align*}
	P(\FP|x \geq t^{*}q) &= (t^* - t_x)/t^*\\
		           &= 1 - \frac{x+q}{u+q}
\end{align*}

If we assume that, within a partition, the distribution of domain size is uniform,
we can evaluate $\NFP_{l,u}$ as follows:
\begin{align*}
	\NFP_{l,u} &= N_{l,u}\cdot \sum_{x=l}^{u-1} (1 - \frac{x+q}{u+q})\frac{1}{u-l} \\
		   &= N_{l,u}\frac{1}{u-l} \cdot\sum_{x=l}^{u-1}(1 - \frac{x+q}{u+q})  \\
		   &= N_{l,u} \cdot\frac{u-l+1}{2(u+q)}\\
		   &\le N_{l,u} \cdot \frac{u-l+1}{2u}
\end{align*}

{\bf Case 2: $t_lq \leq l < t^{*}q$ and $t^{*}q \leq u $}
For domains in the sub-interval $[l, t^{*}q)$, their maximum possible containment scores
do not satisfy the query-given containment threshold, but satisfies the
effective containment threshold. The false positive probability
of these domains becomes
\begin{align*}
	P(\FP|t_lq \leq x < t^{*}q) &= (x/q - t_x)/(x/q) \\
			       &= 1 - t_x\frac{q}{x}\\
			       &= 1 - \frac{x+q}{u+q}\cdot \frac{t^{*}q}{x}
\end{align*}

Assuming uniform distribution of domain size as before, 
the $\NFP_{l,u}$ now can be evaluated as
\begin{align*}
	\NFP_{l,u}= N_{l,u}\frac{1}{u-l}\cdot (\sum_{x=l}^{t^{*}q-1} P(\FP|t_lq \le x < t^{*}q) &\\
	+\sum_{x=t^*q}^{u-1} P(\FP|x \ge t^{*}q ) )&\\ 
	\le N_{l,u}\frac{1}{u-l}\cdot \sum_{x=l}^{u-1} P(\FP| x \ge t^{*}q)  &
\end{align*}
Since $x < t^*q$ in the sub-interval $[l, t^{*}q)$, 
	$P(\FP|t_lq \leq x < t^{*}x) \le P(\FP|x \ge t^{*}q)$. 
We still have the inequality:
\begin{equation*}
\NFP_{l,u} \le  N_{l,u} \cdot \frac{u-l+1}{2u}
\end{equation*}

{\bf Case 3: $l < t_lq$ and $t^*q \le u$}
Now the maximum possible containment scores for domains in the sub-interval
$[l, t_lq)$
are below the effective containment threshold.
For these domains, the false positive probability is zero
\begin{equation*}
	P(\FP|x < t_lq) = 0
\end{equation*}

Following the same steps as before, we can show the inequality still holds.
\begin{align*}
	\NFP_{l,u}= N_{l,u}\frac{1}{u-l}\cdot (
	\sum_{x=l}^{t_lq-1} P(\FP| x < t_lq) &\\
	+\sum_{x=t_lq}^{t^{*}q-1} P(\FP|t_lq \le x < t^{*}q)&\\
	+\sum_{x=t^*q}^{u-1} P(\FP| x \ge t^{*}q) )&\\
	\le N_{l,u}\frac{1}{u-l}\cdot \sum_{x=l}^{u-1} P(\FP|x \ge t^{*}q)  &\\
	\le N_{l,u} \cdot \frac{u-l+1}{2u}&
\end{align*}

{\bf Case 4: $l < t_lq$ and $t_uq \le u < t^*q$}
Since the probabilities of false positives in the intervals 
$[l, t_uq)$ and $[t_uq, u)$ are
both less than $P(\FP|x \ge t^*q)$, the inequality still holds, follow the same reasoning.

{\bf Case 5: $u < t_uq$}
The probability of false positive is 0 in the complete interval $[l,u)$, thus the inequality
still holds.

Thus Proposition 2.

\subsection*{Proof of Theorem 1}

Based on the proof for Proposition~\ref{prop:2} in the previous section,
we can derive that 
the number of false positives in any partition is monotonic with respect
to the width of the partition -
increasing $u$, decreasing $l$, or both results in higher $\NFP_{l,u}$. 

For the case of $n$ partitions, we prove that equi-$\NFP_i$ is an optimal 
partitioning using proof by contradiction. Let $\Pi_1$ be a partitioning 
with $n$ equi-$\NFP_i$ partitions. Since all partitions have equal number 
of false positives, the cost of $\Pi_1$ is
\begin{equation}
	\cost(\Pi_1) = \max_i\NFP_i = \NFP_i
\end{equation}

We prove that $\Pi_1$ is optimal by showing that any change in partitioning will cause 
the partitioning to become non-optimal.
Assume $\Pi_1$ is not optimal, meaning there exists a partitioning $\Pi_2$
with $\cost(\Pi_2) < \cost(\Pi_1)$. Partitioning $\Pi_2$ can be resulted from 
$\Pi_1$ by changing the boundaries of at least two adjacent partitions $p_j$ and
$p_k$. Assume the boundary of $p_j$ and $p_k$ is $u_j$, which is the upper bound of 
$p_j$. Without loss of generality, assume increasing $u_j$ results in the optimal $\Pi_2$.
According to the monotonocity property of $\NFP$ of a partition w.r.t the partition,
increasing $u_j$ implies $\NFP_j > \NFP_i$, where $\NFP_i$ is the number of false
negatives in partition $p_j$ before increasing the upper bound. Since, the increase of $u_j$ 
decreases $\NFP_k$ and increases $\NFP_j$, while $\NFP$ remains unchanged, $\cost(\Pi_2) = \NFP_j$.
We argued earlier that $\NFP_j > \NFP_i$, therefore, $\cost(\Pi_2) > \cost(\Pi_1)$. According to 
the definition of optimal partitioning, $\Pi^* = \argmin_{\Pi}\max_{i}\NFP_i$, the 
partitioning $\Pi_2$ is not optimal. We proved that changing the partitions in an equi-$\NFP_i$
does not result in a lower cost partitioning. Therefore, equi-$\NFP_i$ is optimal.
\qed \\

\subsection*{Asymmetric Minwise Hashing}
\label{sec:asymanalysis}
In this section we provide a breif overview of Asymmetric Minwise Hashing
and the reason why it is not suitable for domain search over very skewed domain size
distribution.

\subsubsection*{Preliminaries}
Asymmetric Minwise Hashing technique uses asymmetric 
transformation (padding) on domains before inserting them into an MinHash LSH index.
Let $M$ be the upper bound of domain size of indexed domains before padding.
Padding is to add new values to domains to make them all have size equals to $M$. 
Since the new values have no overlap with the values of the original domains and 
queries, the containment scores stay the same after padding. 
Thus, we can update Equation~\ref{eq:s-t} for a padded domain by simply replacing 
the domain size $x$ with $M$ to obtain the conversion from containment to 
Jaccard similarity:
\begin{equation}
	\label{eq:s-t-asym}
    	\Hs{M,q}(t) = \frac{t}{\frac{M}{q} + 1 - t}
\end{equation}
In the above equation, the conversion to Jaccard similarity only depends on the
containment score $t$ and the query domain size $q$, since $M$ is a constant.
Furthermore, the converted Jaccard similarity $\Hs{M,q}(t)$ is monotonic with respect
to $t$.

In a MinHash LSH, given a query domain, the probability of an indexed domain becoming 
a candidate is determined by its Jaccard similarity with the query domain
(see Section~\ref{sec:lsh}).
Asymmetric Minwise Hashing uses padded domains to build a MinHash LSH index.
Because of Equation~\ref{eq:s-t-asym}, the Jaccard similarity of an padded domain
is monotonic with respect to its containment score.
Consequently, the probability of an padded indexed domain is also monotonic with 
respect to its containment score.
This implies that the Jaccard similarity near neighbours converges to the containment
near neighbours, making containment search possible.

\subsubsection*{Impact of Skewed Domain Sizes}
We will show with an example, 
how skewed domain size distribution can impact the recall of Asymmetric Minwise Hashing.

\begin{figure}[t]
    \includegraphics[width=\columnwidth,natwidth=1, natheight=1]{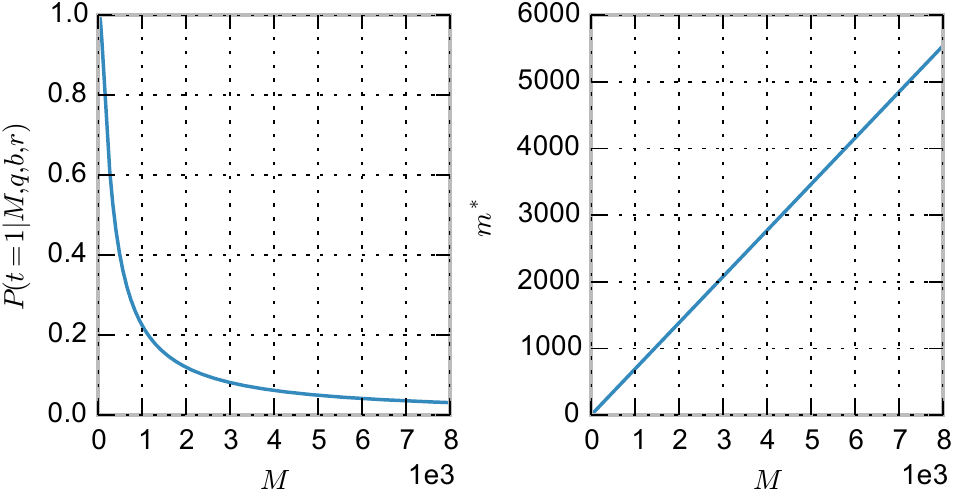}
    \caption{Left: the probability of being selected for a domain with $t = 1.0$;
    	Right: the minimum number of hash functions required to maintain the 
	probability above 0.5}
    \label{fig:asymanalysis}
\end{figure}

Assume a domain has a containment score of 1.0.
Using Equation~\ref{eq:s-t-asym} and \ref{eq:probLSH}, we can derive the probability
of the padded domain being selected as a candidate by MinHash LSH:
\begin{equation}
    \label{eq:prob-asymanalysis}
    P(t=1|M,q,b,r) = 1 - \left(1 - \left(\frac{q}{M}\right)^r\right)^b
\end{equation}
We can assume that we can tune the LSH to maximize the probability by letting 
$r = 1$ and $b = 256$ given the number of hash functions is 256.
We can further assume $q = 1$ without loss of generality.
Figure~\ref{fig:asymanalysis} shows the probability of the domain being selected
decreases very fast as the upper bound of domain size $M$ increases,
even when the MinHash LSH is tuned to maximize the probability.

Thus, when the maximum domain size $M$ is very large relative to the query domain 
size, the probability for qualifying domains (after padding) being selected as 
candidates is almost zero, resulting in very low recall.
This scenario can happen in practice when the domain sizes follow a power-law 
distribution, as in the Canadian Open Data repository and WDC Web Table dataset.
Our experiments confirm this result (see Section~\ref{sec:accuracy}).

We can still maintain the probability at certain level by increasing the number of 
hash functions we use.
The right plot of Figure~\ref{fig:asymanalysis} is based on
Equation~\ref{eq:prob-asymanalysis}. It shows that the minimum number of hash functions
required ($m^*$) to maintain the probability above 0.5 increases linearly with
respect to $M$.
Since the cost of building and searching MinHash LSH index increases with the 
number of hash functions, having too many hash functions will result in low 
performance.

\end{document}